# Radiation rebound and quantum splash in electron-laser collisions

Z. Gong,[1,2] R. H. Hu,[1,3] J. Q. Yu,[1,4] Y. R. Shou,[1] A. V. Arefiev,[5,6] and X. Q. Yan[1,7,*]

[1]SKLNPT, KLHEDP and CAPT, School of Physics, Peking University, Beijing 100871, China
[2]Center for High Energy Density Science, The University of Texas at Austin, Austin, Texas 78712, USA
[3]College of Physics, Sichuan University, Chengdu 610065, China
[4]School of Physics and Electronics, Hunan University, Changsha 410082, China
[5]Department of Mechanical and Aerospace Engineering,
University of California at San Diego, La Jolla, California 92093, USA
[6]Center for Energy Research, University of California at San Diego, La Jolla, California 92093, USA
[7]Collaborative Innovation Center of Extreme Optics, Shanxi University, Taiyuan, Shanxi 030006, China

The radiation reaction (RR) is expected to play a critical role in light-matter interactions at extreme intensity. Utilizing the theoretical analyses and three-dimensional (3D) numerical simulations, we demonstrate that electron reflection, induced by the RR in a head-on collision with an intense laser pulse, can provide pronounced signatures to discern the classical and quantum RR. In the classical regime, there is a precipitous threshold of laser intensity to achieve the whole electron bunch rebound. However, this threshold becomes a gradual transition in the quantum regime, where the electron bunch is quasi-isotropically scattered by the laser pulse and this process resembles a water splash. Leveraged on the derived dependence of classical radiation rebound on the parameters of laser pulses and electron bunches, a practical detecting method is proposed to distinguish the quantum discrete recoil and classical continuous RR force.

The motion of charged particles within an electric-magnetic field is always accompanied by the emission of radiation and its corresponding recoil [1]. Normally, the radiation reaction (RR) effect is extremely small compared to the dominant Lorentz force, but it has a significant impact on violent universe environments, such as the Crab Nebula invoked by a stellar explosion [2,3], curved space-time near magnetized black holes [4,5], and relativistic current sheets in pulsar wind [6]. On Earth, avoiding the energy dissipation of the RR and exploiting its pertinent photon emission, scientists employ linear accelerators to generate high-energy electron beams [7,8] and meanwhile utilize the synchrotron mechanism to produce brilliant x- and gamma-ray light sources [9–14].

Since the seminal theoretical exploration in 1938 when Dirac proposed the renormalized Abraham-Lorentz-Dirac equation [15–17], the debate about the exact form of the RR has never ended [18–22]. Currently, in the classical regime, the prevailing Landau-Lifshitz (LL) force [18], circumventing the notorious runaway solution, describes the self-consistent continuous reaction of electron dynamics. In the realm of quantum electrodynamics (QED), the discrete recoil characterized by $\chi_e = |F_{\mu\nu}p^\nu|/m_e E_s$ [20] is more suitable to reproduce multiphoton scattering [23–29] and quantum quenching [30]. Here, $F_{\mu\nu}$ is the field tensor, $p^\nu$ is the four-vector momentum of the electron (mass $m_e$ and charge $e$), and $E_s = m_e^2 c^3/q_e\hbar \simeq 1.3 \times 10^{18}$ V/m is the Sauter-Schwinger field [31,32].

Through the Lorentz boost to amplify the electric field in the instantaneous frame, the relativistic electron (energy $\varepsilon_e$) colliding with a laser pulse (intensity $I_0$) is the optimum configuration to investigate the RR [33–38], where the invariant parameter is estimated as $\chi_e \approx 0.66\varepsilon_e[\text{GeV}](I_0[10^{23}\text{ W}/\text{cm}^2])^{1/2}$. For the weak quantum situation $\chi_e < 1$, experiments have already elucidated the energy-damping effect of the RR via all-optical Compton backscattering (at $\chi_e \lesssim 0.2$) [39,40]. For the strong quantum regime $\chi_e > 1$, theoretical studies demonstrate that both radiation damping and quantum discrete recoil substantially influence the interaction processes [41–54]. However, there is still no experimental result to examine the quantum stochastic discrete recoil, because, when a pulse with intensity $I_0 \lesssim 10^{21}$ W/cm² counter collides with GeV electrons, the quantum parameter reads merely $\chi_e \lesssim 0.1$. With the development of optics [55–57], the state-of-the-art laser technology [58] provides the next-generation facilities [59–62] with an opportunity to achieve intensities beyond $10^{23}$ W/cm$^2$ (e.g., the extreme light infrastructure project [63,64]), which opens up a new avenue to attain $\chi_e \gtrsim 1$ for the above target.

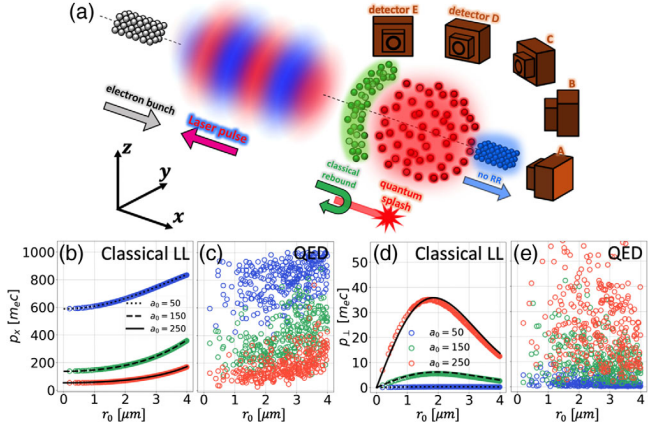

FIG. 1.  (a) The schematic of a collision between the electron bunch and laser pulse. (b) and (c) present the dependence of $p_x$ on the electron's initial transverse coordinate $r_0$ for the classical LL and QED case, respectively. (d) and (e) are the same as (b) and (c) but for $p_\perp$. The black lines in (b) and (d) correspond to the theory of Eqs. (5) and (9).

In this paper, we propose a theoretical scheme to distinguish the continuous damping and stochastic discrete recoil in classical and quantum RR models, where an ultrarelativistic electron bunch from wakefield accelerators [65–68] is used to collide with a counterpropagating high-intensity ($I_0 \gtrsim 10^{23}$ W/cm$^2$) laser pulse shown in Fig. 1(a). When the RR is taken into account, our scheme has two distinct features. First, in the classical regime, the longitudinal damped electrons will be deflected or even permanently reflected by the synergy between transverse expulsion and the insufficient longitudinal compensation of the laser ponderomotive force [69]. Second and more importantly, in the quantum model, the stochastic recoil makes electrons have quasi-isotropic distribution over a wide range of parameters. These phenomena of electron eventual patterns, comparable with the intricate photon radiation spectral [70–72], are more feasible to be detected in an experiment to differentiate the discrete quantum recoil from the classical continuous one. Three models are employed in characterizing the interaction here. For clearness, each of them is given an abbreviation when being discussed in text or exhibited in the figures: first, the Lorentz force without the radiation reaction effect, which is named as *no RR*, second, the classical radiation reaction force of the Landau-Lifshitz formula, which is termed as *classical LL*, and third, the quantum discrete radiation recoil, which is abbreviated as *QED*.

In the head-on collision between an electron and a short pulse, three distinct stages exist. (i) Slow down.—The electron dissipates energy when colliding with the laser pulse. (ii) Reacceleration.—The electron is reaccelerated and begins to surf with the laser wave front, where the energy loss induced by the RR is minimal. (iii) Transverse expulsion.—The transverse ponderomotive force expels the electron while it is still copropagating with the pulse. After leaving the laser pulse, the electron retains its inertial motion, which is termed as the permanent reflection.

To characterize the electron dynamics, we consider that the electron (Lorentz factor $\gamma_0 \gg 1$) initially possesses a positive momentum $p_x$ and a counterpropagating linearly polarized (LP) laser pulse with spot size $\sigma_0$ and $1/e$ half duration $\tau_0$, whose electric and magnetic fields can be expressed as $E_y \propto a_0 \exp\{-r^2/\sigma_0^2\}\exp\{-(\phi-\phi_0)^2/(\omega_0\tau_0)^2\}$ and $B_z = -E_y/c$, where $r = \sqrt{y^2+z^2}$ the transverse coordinate, $\phi = \omega_0 t + kx$ the relative phase and $\phi_0$ the initial condition. Taking account of the relation of $\nabla \cdot \mathbf{E} = 0$ and $\nabla \cdot \mathbf{B} = 0$, the laser longitudinal fields $E_x$ and $B_x$ intrinsically exist. The classical RR model adopts the classical LL force, which is formulated as $\mathbf{f}_{LL} = -(4\pi r_e/3\lambda_0)m_e\omega_0 a_s^2\chi_e^2\mathbf{v}$ with $\lambda_0$ ($\omega_0$) the laser wavelength (frequency), $r_e = e^2/m_e c^2$ the classical electron radius, and $a_s = eE_s/m_e c\omega_0$ the normalized Schwinger field. For simplicity, we introduce the normalization $t \to \omega t$, $x \to kx$, $\mathbf{v} \to \mathbf{v}/c$, $\mathbf{p} \to \mathbf{p}/m_e c$, $\mathbf{E} \to e\mathbf{E}/m_e c\omega_0$, and $\mathbf{B} \to e\mathbf{B}/m_e\omega_0$.

Below, we estimate the eventual momentum value when the electron is merely damped and deflected rather than reflected by the counterpropagating laser pulse. Here, the small deflection approximation (SDA) is assumed when the electron is not reflected, and the electron longitudinal momentum is the dominant term in its relativistic Lorentz factor, i.e., $p_x^2 \gg p_\perp^2 = p_y^2 + p_z^2$. For a relativistic electron, the condition of the SDA can lead to relations $\gamma \approx p_x$, $v_x \approx 1$, and $\chi_e \approx 2p_x a_0/a_s$ as well. The electron longitudinal dynamics under the continuous classical LL force can be characterized by

$$\frac{dp_x}{dt} = -E_x - v_y B_z - \eta v_x \chi_e^2, \qquad (1)$$

$$\frac{d\gamma}{dt} = -v_x E_x - v_y E_y - \eta \chi_e^2, \qquad (2)$$

where the coefficient $\eta = (4\pi r_e/3\lambda_0)m_e\omega_0 a_s^2 \approx 1983$ for $\lambda_0 = 1\ \mu\text{m}$. The sum of Eqs. (1) and (2) gives

$$\frac{dp_x}{dt} + \frac{d\gamma}{dt} = -(1+v_x)(\eta\chi_e^2 + E_x). \qquad (3)$$

The oscillating $E_x$ in the right-hand side of Eq. (3) is negligible when compared with the $\eta\chi_e^2$ in contributing to the damping of the electron longitudinal momentum. Meanwhile, after substituting the relations $\gamma \approx p_x$, $v_x \approx 1$, and $\chi_e \approx 2p_x a_0/a_s$ into Eq. (3), the damping rate of the electron longitudinal momentum is expressed as

$$\frac{dp_x}{dt} \approx -2\eta\chi_e^2. \qquad (4)$$

The integral of Eq. (4) tells us the final longitudinal momentum of the electron $p_{xf}$ after colliding with the pulse:

$$p_{xf} \approx \frac{\gamma_0}{1 + \sqrt{\pi/2}\gamma_0\eta(a_0/a_s)^2\tau_0 \exp(-2r_0^2/\sigma_0^2)}, \quad (5)$$

where the pulse spot size is assumed nearly constant $\sigma \approx \sigma_0$ and the electron transverse movement is assumed to be insignificant in the calculation (i.e., $r \approx r_0$). It is worth pointing out that Eq. (5) will be retrieved to the damped electron longitudinal momentum in the plane wave situation [33] if the laser transverse profile term $\exp(-2r_0^2/\sigma_0^2)$ is neglected.

The equations of electron transverse momentum can be written as

$$\frac{dp_y}{dt} = -E_y - v_z B_x + v_x B_z - \eta v_y \chi_e^2, \quad (6)$$

$$\frac{dp_z}{dt} = v_y B_x - \eta v_z \chi_e^2. \quad (7)$$

Considering the vector potential of laser pulse **A** with relation $-\partial \mathbf{A}/\partial t = \mathbf{E}$ and $\nabla \times \mathbf{A} = \mathbf{B}$, we can replace $dA_y/dt = -E_y + v_x B_z - v_z B_x + v_y \partial A_y/\partial y$ and $B_x = -\partial A_y/\partial z$ into the above transverse dynamics equation and then obtain

$$\frac{d(p_\perp - A_\perp)}{dt} \approx -\frac{1}{2\gamma}\frac{\partial A_\perp^2}{\partial y} - \eta v_\perp \chi_e^2, \quad (8)$$

where $p_y \approx A_y$ is utilized for the first-order approximation. As the second term on the right-hand side of Eq. (8) is oscillating and $v_\perp \approx 0$, $\eta v_\perp \chi_e^2$ is neglected in estimating the final net transverse momentum. The first term on the right-hand side, $-1/(2\gamma)\partial A_\perp^2/\partial y$, is the transverse ponderomotive force of the laser pulse. When the electron dumps more energy, the transverse expulsion becomes more considerable. Substituting the $p_{xf}(t) \approx \gamma(t)$ into Eq. (8), the final transverse momentum is approximated as

$$p_{\perp f} \approx \frac{a_s^2}{2\eta}\frac{\varpi}{\gamma_0}(1 + \rho\gamma_0\varpi)\frac{r_0}{\sigma_0^2}, \quad (9)$$

where $\varpi = \sqrt{\pi/2}\eta(a_0/a_s)^2\tau_0 \exp(-2r_0^2/\sigma_0^2)$ and $\rho \approx 0.33$ is a constant accounting for the effective Lorentz factor involved in the transverse ponderomotive force. The net contribution of $p_{\perp f}$ predominantly comes from the longitudinal damping and subsequent transverse expulsion. When the transverse momentum becomes comparable with the damped longitudinal one, the SDA is no longer valid, which is considered as the criterion of permanent reflection induced by the classical RR. From $p_{\perp f} \sim p_{xf}$, the threshold is estimated as

$$a_*^2 \approx \frac{1.7 a_s^{4/3}}{\eta^{2/3}\tau_0}\left(\frac{\sigma_0^2}{r_0}\right)^{1/3} \exp(2r_0^2/\sigma_0^2). \quad (10)$$

Here, $a_*$ is the required amplitude to trigger the permanent reflection for a single electron. In order to estimate the reflection criterion for the whole electron bunch, we leverage on the averaged value of threshold $a_*$ over all the single electrons. Correspondingly, the threshold of rebounding the whole bunch of electrons can be approximated via the average of $a_*$ as

$$a_{th} \approx \frac{0.4 a_s^{2/3} \sigma_0^{1/3}}{\eta^{1/3}\tau_0^{1/2} r_0^2} \iint r'^{-1/12} \exp\frac{r'^2}{\sigma_0^2} r d\psi dr, \quad (11)$$

where $r' = d_0^2 + r^2 + 2d_0 r \cos\psi$ and $d_0$ is the misalignment between the central axis of the electron bunch and the colliding pulse. At the field strength $a_{th}$, around 50% of electrons of the bunch are reflected. The number is not exactly 50%, because the average weight is not a uniform function. For a specific case $d_0 = 0$, Eq. (11) can be analytically calculated as $a_{th,d=0} \approx \text{Re}\{\mathcal{C}(-1)^{1/12}[\Gamma(\frac{11}{12}, -\frac{r_0^2}{\sigma_0^2}) - \Gamma(\frac{11}{12})]\}$, where Re denotes the real part, $\Gamma(l, x)$ is the incomplete gamma function, and the constant factor $\mathcal{C} = 1.2 a_s^{2/3}\sigma_0^{13/6}/(\eta^{1/3}\tau_0^{1/2}r_0^2)$.

To prove the above analyses, simplified 3D simulations are performed, where the model similar to Refs. [73–77] is used in the numerical algorithm (Sec. S1 in Supplemental Material [78]). The simulation box has a size of 100 μm × 100 μm × 100 μm, and the LP pulse has the waist radius $\sigma_0 = 5$ μm, peak amplitude $50 \leq a_0 \leq 450$, temporal duration $\tau_0 = 3T_0$ ($T_0 \approx 3.3$ fs), and wavelength $\lambda_0 = 1$ μm. It should be noted that the laser amplitude $a_0 \gg 1$ accords with the local constant field approximation (LCFA) [79–83] and, thus, issues of quantum interference [84] are insignificant. A bunch of electrons, represented by $N = 10^5$ macroparticles with initial $\gamma_0 = 1000$ and uniformly distributed within a cylinder with length 5 μm and radius 2.5 μm, are injected from the left side to counter collide with the pulse. Since the electron bunch from wakefield accelerators is underdense and ultrarelativistic, the charge separation force of electrons is negligible [85]. Moreover, the practical energy spread $\delta\gamma_{\text{FWHM}}/\gamma_0 = 5\%$, beam divergence 5 mrad, and misalignment $d_0$ are considered when simulating the interaction processes in Figs. 3 and 4.

The dependence of the final electron momentum on its initial coordinates $r_0$ is illustrated in Figs. 1(b) and 1(d), where the simulations are in good agreement with the theory of $p_x = 1000/[1 + 2.79(a_0[100])^2 \exp(-2r_0^2/\sigma_0^2)]$ from Eq. (5) and $p_\perp = 18.8(a_0[100])^2 \exp(-2r_0^2/\sigma_0^2)[1 + 0.93(a_0[100])^2 \exp(-2r_0^2/\sigma_0^2)](r_0/\sigma_0^2)$ from Eq. (9) when $a_0 = 50$, 150, and 250. For $a_0 \leq 250$, the relatively small transverse $p_y$ guarantees the validity of the SDA, but the SDA is no longer valid for a strong laser pulse with $a_0 \geq 300$. The evolution of $p_x$ and the trajectories of four typical electrons at $a_0 = 350$ are plotted in Fig. 2(a). All of them experience a drastic energy damping during $49T_0 < t < 51T_0$ as $R_c = \alpha\chi_e a_0 \approx 4.36$ manifests that the average energy radiated by the electron in one laser period is comparable to its initial energy [23,41,42,86]. Here, $\alpha \approx 1/137$ is the fine structure constant. Electrons no. (1)–(3) undergo a temporal reflection [70] and are reaccelerated during $51T_0 < t < 60T_0$. For electron no. (1) with $r_0 = 0.0$ $\mu$m, it cannot obtain a net transverse expulsion, so that the tail down ramp of the laser pulse eventually pushes it back to a positive $p_x = 28.5$, which agrees well with $p_{xf}|_{a_0=350, r_0=0} = 28.3$ predicted by Eq. (5). The transverse expulsion makes electrons no. (2) and (3) slide out of the central axis, and, subsequently, they do not have a chance to witness the down ramp of the laser pulse, whence a permanent reflection occurs. The radiation damping for electron no. (4) is not enough to trigger the temporal reflection, and, thus, it is only deflected by the ponderomotive force. The simulation threshold of permanent reflection for the classical LL model is shown in Fig. 2(c), where theoretical $a_*$ of Eq. (10) and the calculation for the QED model are included as well. The parameters of four electrons in Fig. 2(a) are also marked in Fig. 2(c).

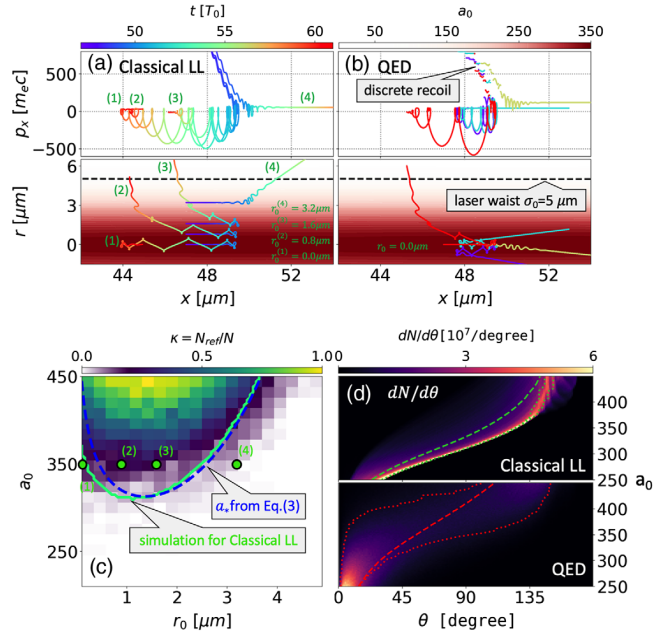

FIG. 2. (a) The evolution of $p_x$ versus $x$ and trajectories of four electrons ($r_0 = 0.0$, 0.8, 1.6, and 3.2 $\mu$m, respectively) characterized by the classical LL for $a_0 = 350$, where the rainbow color scale denotes the time and the white and red color map corresponds to the strength of the laser field amplitude. (b) is the same as (a) but for the QED electrons with the identical $r_0 = 0$ in different colors. (c) The green solid line denotes the $a_*$ simulated from the classical LL model, the blue dashed line refers to theoretical $a_*$ of Eq. (10), and the viridis color map exhibits the calculation of the QED model. (d) The electron angular distribution $dN/d\theta$ for $250 < a_0 < 450$. The dashed green (red) line denotes the average value $\bar{\theta}$ versus $a_0$ for the LL (QED) case, while the dotted line denotes the position of the FWHM.

In the quantum regime, the differential rate of emitting photons (wave vector $k^\nu$) with parameter $\chi_\gamma = |F_{\mu\nu} k^\nu|/E_s$ by an electron with $\chi_e$ reads $d^2 N(\chi_e, \chi_\gamma)/d\chi_\gamma dt = (\sqrt{3}\alpha/2\pi\tau_c)[\chi_e F(\chi_e, \chi_\gamma)]/(\gamma\chi_\gamma)$, where $\tau_c \approx 1.28 \times 10^{-6}$ fs is the Compton time and the quantum synchrotron spectrum $F(\chi_e, \chi_\gamma)$ is validated under the LCFA [87]. The randomness and uncertainty of photon emission manifest that the discrete quantum recoil is like a virtual inelastic collision resulting in the transition from classical radiation cooling to stochastic heating [88–90]. Considering the analytic irreducibility of the QED model, we appeal to the numerical method to investigate the underlying physics. The momentum distributions of the QED electron in Figs. 1(c) and 1(e) tend to be more insensitive and stochastic. The average insensitivity is because the equivalent damping term $g(\chi_e) \mathbf{f}_{LL}$ is modified by a weight factor $g(\chi_e) = (3.7\chi_e^3 + 31\chi_e^2 + 12\chi_e + 1)^{-4/9} < 1$ [35,73], which has already been proved by experiments [39,40,91]. The evolution of $p_x$ and trajectories for four QED electrons (with the identical initial coordinate $r_0 = 0$) is shown in Fig. 2(b), where two random processes are intrinsically incorporated in the discrete quantum emission. One is the possibility of photon generation, and the other is the random value of the emitted energy. The red electron, undergoing sufficient recoil effect during 48 $\mu$m $\leq x \leq$ 49.5 $\mu$m, is easily reflected by the counterpropagating pulse. The yellow electron, experiencing relatively inadequate radiation damping during 48 $\mu$m $\leq x \leq$ 50 $\mu$m, passes through the whole pulse with merely a slight transverse deflection. The electrons marked in cyan and purple are the intermediate case, where they undergo temporal reflection but possess a positive $p_x$ eventually.

The definition of electron reflection can be set as $p_x < 0$, i.e., the angle $\theta = \arctan(p_\perp/p_x) > 90°$. The angular distributions of the electron number for $250 \leq a_0 \leq 450$ are shown in Fig. 2(d). In the upper panel of the classical LL case, the average polar angle $\bar{\theta} = \sum_{i=1}^N \theta_i/N$ increases with the rising of the laser amplitude $a_0$, and the full width at half maximum (FWHM) bandwidth $\theta_b \lesssim 5°$ is quite narrow. Therefore, the reflection of electrons is sensitive to the variation of laser amplitude $a_0$. In contrast, due to quantum weakening effects, the $\bar{\theta}$ of the QED model is a little smaller compared to that of the classical LL case. In the lower panel in Fig. 2(d), the bandwidth $\theta_b > 15°$ of the QED case is much wider than the LL one due to its stochastic emission processes, and it even exhibits a quasi-isotropic feature around $350 < a_0 < 400$.

Because of the difficulty of tracking electrons in experiments, measuring the reflection efficiency $\kappa$ is more feasible, where $\kappa = N_{\text{ref}}/N$ is the ratio of the number of reflected electron $N_{\text{ref}}$ to the whole bunch $N$. The parameter scans of $\kappa$ with laser amplitude $210 \leq a_0 \leq 450$, spot size $3 \leq \sigma_0[\mu m] \leq 9$, duration $1.5 \leq \tau_0[T_0] \leq 5.5$, initial energy $500 \leq \gamma_0 \leq 2000$, and misalignment $0 \leq d_0[\mu m] \leq 2$, are shown in Fig. 3. There is a pronounced separatrix in Figs. 3(a), 3(c), 3(e), and 3(g) between $\kappa = 0$ and 1.0, where the thresholds of Eq. (11) drawn in black dashed lines correspond well with the 3D simulations. In Figs. 3(b), 3(d), 3(f), and 3(h), the sharp transition becomes a gradual up ramp for the QED case because of the stochastic RR and uncertain trajectory. In Figs. 3(a) and 3(b), the required amplitude is increased when the laser spot size $\sigma_0$ varies from 4.8 to 9.0 μm, due to the weaker transverse ponderomotive expulsion induced by a larger spot size. The abnormal result for 3 μm $< \sigma_0 <$ 4.8 μm is because the spot size is too small to interact with most of the electrons and the edge parts are hardly reflected. In Figs. 3(c) and 3(d), there is a complementary relationship between the laser temporal duration $\tau_0$ and peak amplitude $a_0$. Figure 3(e) shows that the reflection threshold is not sensitive to the initial electron energy $\gamma_0$ for the classical case, whereas bigger amplitude $a_0$ is required to splash the electron bunch with the larger $\gamma_0$ in Fig. 3(f). The distinct pattern between the classical LL and QED models still exists when the misalignment is relatively small at $d_0 \leq 1$ μm as shown in Figs. 3(g) and 3(h). The same parameter scans without considering the energy spread and beam divergence are also performed, and the physics does not change (Sec. S2 in Supplemental Material [78]).

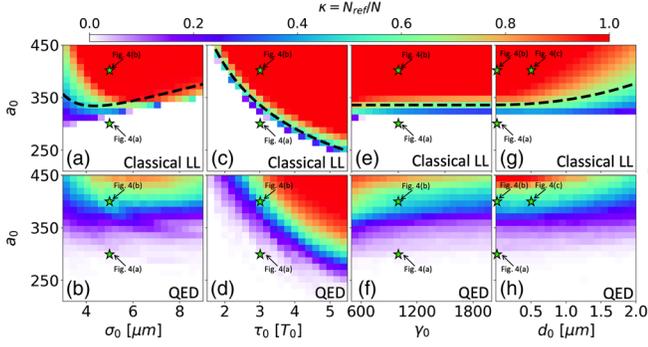

FIG. 3. (a),(b) The ratio $\kappa$ obtained from 3D simulations figured in $a_0 - \sigma_0$ space for $\tau_0 = 3T_0, \gamma_0 = 1000$, and $d_0 = 0$ μm. (c),(d) and (e),(f) are identical with (a),(b), except the parameter planes are in $\tau_0 - a_0$ (for $\sigma_0 = 5$ μm, $\gamma_0 = 1000$, and $d_0 = 0$ μm) and $\gamma_0 - a_0$ (for $r_0 = 5$ μm, $\tau_0 = 3T_0$, and $d_0 = 0$ μm), respectively. (g),(h) exhibit the results in $a_0 - d_0$ space, where $\sigma_0 = 5$ μm, $\tau_0 = 3T_0$, and $\gamma_0 = 1000$. The black dashed lines in the classical LL case denote the theoretical threshold of Eq. (11). The green pentagrams correspond to the parameters used in Figs. 4(a)–4(c).

Exploiting the threshold in Eq. (11), we propose to use imaging plates (IPs) [92–94] to record the electron number in a specific direction and, thus, determine which RR model is closer to the exact form. All the detectors, as illustrated in the schematic (Fig. 1), are set parallel with the plane $z = 0$ and have the azimuthal angles $\theta_{xy}$ in the $x$–$y$ plane: 0° (A), 30° (B), 60° (C), 90° (D), and 120° (E). Assuming the detector plates with an area of 20 cm × 20 cm and 50 cm from the interaction region, we find that each detector can collect the electron within a conical angle of $\theta_c \approx 11°$ along its direction. The electron spatial distributions at time $t = 80T_0$ for (a) $a_0 = 300$ and $d_0 = 0$, (b) $a_0 = 400$ and $d_0 = 0$, and (c) $a_0 = 400$ and $d_0 = 0.5$ μm are shown in Fig. 4, where the parameters ($\gamma_0 = 1000$, $\sigma_0 = 5$ μm, and $\tau_0 = 3T_0$) are selected as the pentagrams marked in Fig. 3. In the $a_0 = 400$ case [Fig. 4(b)], the QED electrons behave like the isotropic water splash, while the classical LL electrons are almost completely reflected with $\theta > 90°$. The electron number collected by each of the detectors is plotted in the upper panel in Fig. 4(d), where the charge of the whole electron bunch is assumed as 160 pc. The distinct difference indicates that, if the electrons are exclusively collected by detector E, the continuous classical LL force is prioritized; otherwise, the QED discrete recoil tends to be more accurate. Furthermore, a slight misalignment $d_0 \neq 0$ [Fig. 4(c)] does not result in a pronounced difference in the electron distribution pattern compared with the perfect aligned configuration [Fig. 4(b)], and merely a small fraction of electrons are recorded by detector D [middle panel in Fig. 4(c)]. In Eq. (11), increasing the duration $\tau_0$ is equivalent to reducing the required $a_{th}$. The result of an alternative parameter $a_0 = 300$ and $\tau_0 = 6T_0$ [lower panel in Fig. 4(d)] also exhibits the capability to distinguish these two RR models.

In conclusion, we demonstrate that the electron distribution pattern induced by the RR when counter colliding with a strong laser pulse is different between the models of classical continuous damping and quantum discrete recoil. In the classical RR, there is a precipitous threshold determining whether the electron bunch is reflected or not. Instead, the threshold becomes a smooth transition for the quantum RR. Leveraging on the theoretical derivation and numerical simulations, a practical scheme is proposed to measure the electron number in specific directions to determine the exact form of the RR, which is instructional for future experiments.

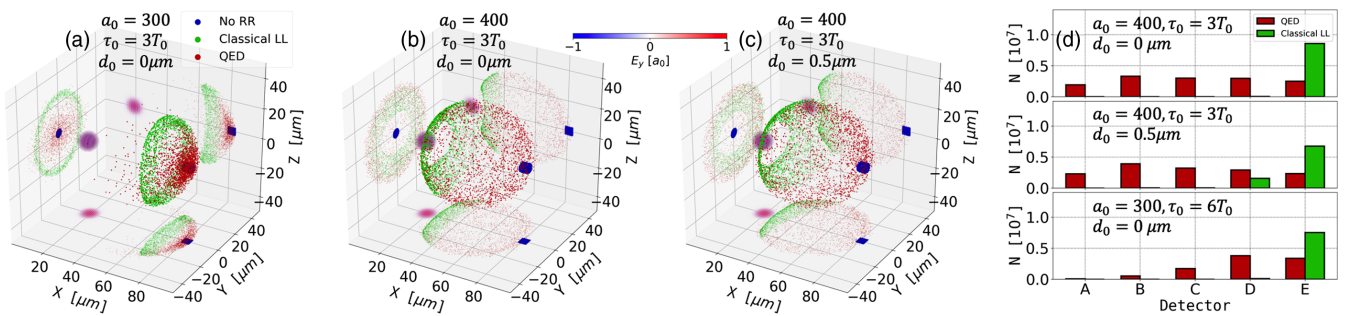

FIG. 4. (a)–(c) The electron three-dimensional distributions at $80T_0$ for different parameters (see Supplemental Material [78]). The blue, red, and green dots indicate the electron dynamics characterized by the Lorentz equation (without the RR), classical LL force, and QED discrete recoil, respectively. (d) presents the number of electrons measured by each detector.

This work has been supported by National Natural Science Foundation of China (Grants No. 11921006 and No. 11535001). This research was supported by NSF (Grants No. 1632777 and No. 1821944). This work was in part funded by the United Kingdom EPSRC Grants No. EP/G054950/1, No. EP/G056803/1, No. EP/G055165/1, and No. EP/M022463/1. This work was supported in part by the scholarship from China Scholarship Council (CSC) under Grant CSC No. 201706010038. J. Q. Y. acknowledges the Fundamental Research Funds for the Central Universities. Z. G. thanks Felix Mackenroth for useful discussion. The simulations were carried out by using the allocation in Texas Advenced Computing Center (TACC) and Shanghai Super Computation Center.

[1] J. D. Jackson, *Electrodynamics* (Wiley Online Library, New York, 1975).

[2] M. Tavani, A. Bulgarelli, V. Vittorini, A. Pellizzoni, E. Striani, P. Caraveo, M. Weisskopf, A. Tennant, G. Pucella, A. Trois *et al.*, Discovery of powerful gamma-ray flares from the crab nebula, Science **331**, 736 (2011).

[3] B. Cerutti, G. R. Werner, D. A. Uzdensky, and M. C. Begelman, Three-dimensional relativistic pair plasma reconnection with radiative feedback in the crab nebula, Astrophys. J. **782**, 104 (2014).

[4] A. Aliev and D. Gal'Tsov, "Magnetized" black holes, Sov. Phys. Usp. **32**, 75 (1989).

[5] A. Tursunov, M. Kološ, Z. Stuchlík, and D. V. Galtsov, Radiation reaction of charged particles orbiting a magnetized schwarzschild black hole, Astrophys. J. **861**, 2 (2018).

[6] C. H. Jaroschek and M. Hoshino, Radiation-Dominated Relativistic Current Sheets, Phys. Rev. Lett. **103**, 075002 (2009).

[7] H. Wiedemann, *Particle Accelerator Physics* (Springer, New York, 2015).

[8] M. Chodorow, E. Ginzton, W. Hansen, R. Kyhl, R. Neal, W. Panofsky, and Staff, Stanford High-Energy Linear Electron Accelerator (Mark III), Rev. Sci. Instrum. **26**, 134 (1955).

[9] H. Winick and S. Doniach, *Synchrotron Radiation Research* (Springer Science & Business, New York, 2012).

[10] Synchrotron light sources of the world, https://lightsources.org/lightsources-of-the-world/.

[11] S. Corde, K. T. Phuoc, G. Lambert, R. Fitour, V. Malka, A. Rousse, A. Beck, and E. Lefebvre, Femtosecond x rays from laser-plasma accelerators, Rev. Mod. Phys. **85**, 1 (2013).

[12] F. Albert and A. G. Thomas, Applications of laser wakefield accelerator-based light sources, Plasma Phys. Controlled Fusion **58**, 103001 (2016).

[13] T. Huang, A. Robinson, C. Zhou, B. Qiao, B. Liu, S. Ruan, X. He, and P. Norreys, Characteristics of betatron radiation from direct-laser-accelerated electrons, Phys. Rev. E **93**, 063203 (2016).

[14] M. Li, L. Chen, D. Li, K. Huang, Y. Li, Y. Ma, W. Yan, M. Tao, J. Tan, Z. Sheng *et al.*, Collimated gamma rays from laser wakefield accelerated electrons, Matter Radiat. Extremes **3**, 188 (2018).

[15] M. Abraham and A. Föppl, *Theorie der Elektrizität* (BG Teubner, Leipzig, 1905), Vol. 2.

[16] H. A. Lorentz, *The Theory of Electrons and Its Applications to the Phenomena of Light and Radiant Heart: A Course of Lectures Delivered in Columbia University, New York in March and April, 1906* (Teubner, Leipzig, 1916), Vol. 29.

[17] P. A. M. Dirac, Classical theory of radiating electrons, Proc. R. Soc. A **167**, 148 (1938).

[18] L. D. Landau and E. M. Lifshitz, *The Classical Theory of Fields* (Pergamon, New York, 1971).

[19] V. B. Berestetskii, E. M. Lifshitz, and L. Pitaevskii, *Quantum Electrodynamics* (Butterworth-Heinemann, London, 1982), Vol. 4.

[20] V. Ritus, Quantum effects of the interaction of elementary particles with an intense electromagnetic field, J. Russ. Laser Res. **6**, 497 (1985).

[21] I. V. Sokolov, N. M. Naumova, J. A. Nees, G. A. Mourou, and V. P. Yanovsky, Dynamics of emitting electrons in strong laser fields, Phys. Plasmas **16**, 093115 (2009).

[22] S. V. Bulanov, T. Z. Esirkepov, M. Kando, J. K. Koga, and S. S. Bulanov, Lorentz-Abraham-Dirac versus Landau-Lifshitz radiation friction force in the ultrarelativistic electron interaction with electromagnetic wave (exact solutions), Phys. Rev. E **84**, 056605 (2011).

[23] A. Di Piazza, K. Hatsagortsyan, and C. H. Keitel, Quantum Radiation Reaction Effects in Multiphoton Compton Scattering, Phys. Rev. Lett. **105**, 220403 (2010).

[24] F. Mackenroth and A. Di Piazza, Nonlinear Compton scattering in ultrashort laser pulses, Phys. Rev. A **83**, 032106 (2011).

[25] F. Mackenroth and A. Di Piazza, Nonlinear Double Compton Scattering in the Ultrarelativistic Quantum Regime, Phys. Rev. Lett. **110**, 070402 (2013).

[26] Z. Gong, R. H. Hu, Y. R. Shou, B. Qiao, C. E. Chen, X. T. He, S. S. Bulanov, T. Z. Esirkepov, S. V. Bulanov, and X. Q. Yan, High-efficiency $\gamma - ray$ flash generation via multiple-laser scattering in ponderomotive potential well, Phys. Rev. E **95**, 013210 (2017).

[27] A. Gonoskov, A. Bashinov, S. Bastrakov, E. Efimenko, A. Ilderton, A. Kim, M. Marklund, I. Meyerov, A. Muraviev, and A. Sergeev, Ultrabright GeV Photon Source via Controlled Electromagnetic Cascades in Laser-Dipole Waves, Phys. Rev. X **7**, 041003 (2017).

[28] J. Magnusson, A. Gonoskov, M. Marklund, T. Z. Esirkepov, J. Koga, K. Kondo, M. Kando, S. Bulanov, G. Korn, and S. Bulanov, Laser-Particle Collider for Multi-GeV Photon Production, Phys. Rev. Lett. **122**, 254801 (2019).

[29] J. Magnusson, A. Gonoskov, M. Marklund, T. Z. Esirkepov, J. Koga, K. Kondo, M. Kando, S. Bulanov, G. Korn, C. Geddes *et al.*, Multiple-colliding laser pulses as a basis for studying high-field high-energy physics, arXiv:1906.05235.

[30] C. Harvey, A. Gonoskov, A. Ilderton, and M. Marklund, Quantum Quenching of Radiation Losses in Short Laser Pulses, Phys. Rev. Lett. **118**, 105004 (2017).

[31] F. Sauter, Über das Verhalten eines Elektrons im homogenen elektrischen Feld nach der relativistischen Theorie Diracs, Z. Phys. 69, 742 (1931).

[32] J. Schwinger, On gauge invariance and vacuum polarization, Phys. Rev. 82, 664 (1951).

[33] A. D. Piazza, Exact solution of the landau-lifshitz equation in a plane wave, Lett. Math. Phys. 83, 305 (2008).

[34] A. Di Piazza, C. Müller, K. Hatsagortsyan, and C. Keitel, Extremely high-intensity laser interactions with fundamental quantum systems, Rev. Mod. Phys. 84, 1177 (2012).

[35] A. Thomas, C. Ridgers, S. Bulanov, B. Griffin, and S. Mangles, Strong Radiation-Damping Effects in a Gamma-Ray Source Generated by the Interaction of a High-Intensity Laser with a Wakefield-Accelerated Electron Beam, Phys. Rev. X 2, 041004 (2012).

[36] M. Vranic, J. L. Martins, J. Vieira, R. A. Fonseca, and L. O. Silva, All-Optical Radiation Reaction at $10^{21}$ W/cm$^2$, Phys. Rev. Lett. 113, 134801 (2014).

[37] M. Lobet, X. Davoine, E. dHumières, and L. Gremillet, Generation of high-energy electron-positron pairs in the collision of a laser-accelerated electron beam with a multipetawatt laser, Phys. Rev. Accel. Beams 20, 043401 (2017).

[38] X. Geng, L. Ji, B. Shen, B. Feng, Z. Guo, Q. Yu, L. Zhang, and Z. Xu, Quantum reflection above the classical radiation-reaction barrier in the quantum electro-dynamics regime, Commun. Phys. 2, 66 (2019).

[39] K. Poder, M. Tamburini, G. Sarri, A. Di Piazza, S. Kuschel, C. Baird, K. Behm, S. Bohlen, J. Cole, D. Corvan et al., Experimental Signatures of the Quantum Nature of Radiation Reaction in the Field of an Ultraintense Laser, Phys. Rev. X 8, 031004 (2018).

[40] J. Cole, K. Behm, E. Gerstmayr, T. Blackburn, J. Wood, C. Baird, M. J. Duff, C. Harvey, A. Ilderton, A. Joglekar et al., Experimental Evidence of Radiation Reaction in the Collision of a High-Intensity Laser Pulse with a Laser-Wakefield Accelerated Electron Beam, Phys. Rev. X 8, 011020 (2018).

[41] S. Bulanov, T. Z. Esirkepov, Y. Hayashi, M. Kando, H. Kiriyama, J. Koga, K. Kondo, H. Kotaki, A. Pirozhkov, S. Bulanov et al., On the design of experiments for the study of extreme field limits in the interaction of laser with ultrarelativistic electron beam, Nucl. Instrum. Methods Phys. Res., Sect. A 660, 31 (2011).

[42] S. Bulanov, T. Z. Esirkepov, M. Kando, J. Koga, K. Kondo, and G. Korn, On the problems of relativistic laboratory astrophysics and fundamental physics with super powerful lasers, Plasma Phys. Rep. 41, 1 (2015).

[43] A. Gonoskov, A. Bashinov, I. Gonoskov, C. Harvey, A. Ilderton, A. Kim, M. Marklund, G. Mourou, and A. Sergeev, Anomalous Radiative Trapping in Laser Fields of Extreme Intensity, Phys. Rev. Lett. 113, 014801 (2014).

[44] A. Mironov, N. Narozhny, and A. Fedotov, Collapse and revival of electromagnetic cascades in focused intense laser pulses, Phys. Lett. A 378, 3254 (2014).

[45] D. Green and C. Harvey, Transverse Spreading of Electrons in High-Intensity Laser Fields, Phys. Rev. Lett. 112, 164801 (2014).

[46] L. Ji, A. Pukhov, I. Y. Kostyukov, B. Shen, and K. Akli, Radiation-Reaction Trapping of Electrons in Extreme Laser Fields, Phys. Rev. Lett. 112, 145003 (2014).

[47] T. Blackburn, C. Ridgers, J. G. Kirk, and A. Bell, Quantum Radiation Reaction in Laser–Electron-Beam Collisions, Phys. Rev. Lett. 112, 015001 (2014).

[48] H. Wang, X. Yan, and M. Zepf, Signatures of quantum radiation reaction in laser-electron-beam collisions, Phys. Plasmas 22, 093103 (2015).

[49] X.-L. Zhu, T.-P. Yu, Z.-M. Sheng, Y. Yin, I. C. E. Turcu, and A. Pukhov, Dense GeV electron–positron pairs generated by lasers in near-critical-density plasmas, Nat. Commun. 7, 13686 (2016).

[50] S. V. Bulanov, T. Z. Esirkepov, J. Koga, S. Bulanov, Z. Gong, X. Yan, and M. Kando, Charged particle dynamics in multiple colliding electromagnetic waves. Survey of random walk, Lévy flights, limit circles, attractors and structurally determinate patterns, J. Plasma Phys. 83, 2, (2017).

[51] C. P. Ridgers, T. Blackburn, D. Del Sorbo, L. Bradley, C. Slade-Lowther, C. Baird, S. Mangles, P. McKenna, M. Marklund, C. Murphy et al., Signatures of quantum effects on radiation reaction in laser–electron-beam collisions, J. Plasma Phys. 83, 5 (2017).

[52] X. B. Li, B. Qiao, H. X. Chang, H. He, W. P. Yao, X. F. Shen, J. Wang, Y. Xie, C. L. Zhong, C. T. Zhou et al., Identifying the quantum radiation reaction by using colliding ultraintense lasers in gases, Phys. Rev. A 98, 052119 (2018).

[53] B. S. Xie, Z. L. Li, and S. Tang, Electron-positron pair production in ultrastrong laser fields, Matter Radiat. Extremes 2, 225 (2018).

[54] X.-L. Zhu, M. Chen, T.-P. Yu, S.-M. Weng, F. He, and Z.-M. Sheng, Collimated GeV attosecond electron–positron bunches from a plasma channel driven by 10 PW lasers, Matter Radiat. Extremes 4, 014401 (2019).

[55] M. D. Perry and G. Mourou, Terawatt to Petawatt Subpicosecond Lasers, Science 264, 917 (1994).

[56] G. A. Mourou, T. Tajima, and S. V. Bulanov, Optics in the relativistic regime, Rev. Mod. Phys. 78, 309 (2006).

[57] A. Ashkin, G. Mourou, and D. Strickland, The 2018 Nobel Prize in Physics: A gripping and extremely exciting tale of light, Curr. Sci. 115, 1844 (2018).

[58] D. Strickland and G. Mourou, Compression of amplified chirped optical pulses, Opt. Commun. 55, 447 (1985).

[59] J. Zou, C. Le Blanc, D. Papadopoulos, G. Chériaux, P. Georges, G. Mennerat, F. Druon, L. Lecherbourg, A. Pellegrina, and P. Ramirez, Design and current progress of the Apollon 10 PW project, High Power Laser Sci. Eng. 3, e2 (2015).

[60] W. P. Leemans, R. Duarte, E. Esarey, S. Fournier, C. G. R. Geddes, D. Lockhart, C. B. Schroeder, C. Toth, J. Vay, and S. Zimmermann, The berkeley lab laser accelerator (BELLA): A 10 GeV laser plasma accelerator, AIP Conf. Proc. 1299, 3 (2010).

[61] C. Danson, P. Brummitt, R. Clarke, J. Collier, B. Fell, A. Frackiewicz, S. Hancock, S. Hawkes, C. Hernandez-Gomez, P. Holligan et al., Vulcan Petawatt—an ultra-high-intensity interaction facility, Nucl. Fusion 44, S239 (2004).

[62] Exawatt center for extreme light studies, http://www.xcels.iapras.ru.

[63] Extreme light infrastructure project, http://www.eli-laser.eu.
[64] S. Weber, S. Bechet, S. Borneis, L. Brabec, M. Bučka, E. Chacon-Golcher, M. Ciappina, M. DeMarco, A. Fajstavr, K. Falk et al., P3: An installation for high-energy density plasma physics and ultra-high intensity laser–matter interaction at ELI-Beamlines, Matter Radiat. Extremes 2, 149 (2018).
[65] T. Tajima and J. Dawson, Laser Electron Accelerator, Phys. Rev. Lett. 43, 267 (1979).
[66] E. Esarey, C. Schroeder, and W. Leemans, Physics of laser-driven plasma-based electron accelerators, Rev. Mod. Phys. 81, 1229 (2009).
[67] W. P. Leemans, A. J. Gonsalves, H.-S. Mao, K. Nakamura, C. Benedetti, C. B. Schroeder, C. Tóth, J. Daniels, D. E. Mittelberger, S. S. Bulanov et al., Multi-GeV Electron Beams from Capillary-Discharge-Guided Subpetawatt Laser Pulses in the Self-Trapping Regime, Phys. Rev. Lett. 113, 245002 (2014).
[68] A. Gonsalves, K. Nakamura, J. Daniels, C. Benedetti, C. Pieronek, T. de Raadt, S. Steinke, J. Bin, S. Bulanov, J. van Tilborg et al., Petawatt Laser Guiding and Electron Beam Acceleration to 8 GeV in a Laser-Heated Capillary Discharge Waveguide, Phys. Rev. Lett. 122, 084801 (2019).
[69] P. Gibbon, Short Pulse Laser Interactions with Matter (World Scientific, Singapore, 2004).
[70] A. Di Piazza, K. Hatsagortsyan, and C. Keitel, Strong Signatures of Radiation Reaction below the Radiation-Dominated Regime, Phys. Rev. Lett. 102, 254802 (2009).
[71] J.-X. Li, K. Z. Hatsagortsyan, and C. H. Keitel, Robust Signatures of Quantum Radiation Reaction in Focused Ultrashort Laser Pulses, Phys. Rev. Lett. 113, 044801 (2014).
[72] J.-X. Li, Y.-Y. Chen, K. Z. Hatsagortsyan, and C. H. Keitel, Angle-resolved stochastic photon emission in the quantum radiation-dominated regime, Sci. Rep. 7, 11556 (2017).
[73] R. Duclous, J. G. Kirk, and A. Bell, Monte Carlo calculations of pair production in high-intensity laser–plasma interactions, Plasma Phys. Controlled Fusion 53, 015009 (2011).
[74] C. Ridgers, J. G. Kirk, R. Duclous, T. Blackburn, C. Brady, K. Bennett, T. Arber, and A. Bell, Modelling gamma-ray photon emission and pair production in high-intensity laser–matter interactions, J. Comput. Phys. 260, 273 (2014).
[75] A. Gonoskov, S. Bastrakov, E. Efimenko, A. Ilderton, M. Marklund, I. Meyerov, A. Muraviev, A. Sergeev, I. Surmin, and E. Wallin, Extended particle-in-cell schemes for physics in ultrastrong laser fields: Review and developments, Phys. Rev. E 92, 023305 (2015).
[76] M. Vranic, J. L. Martins, R. A. Fonseca, and L. O. Silva, Classical radiation reaction in particle-in-cell simulations, Comput. Phys. Commun. 204, 141 (2016).
[77] Z. Gong, R. Hu, Y. Shou, B. Qiao, C. Chen, F. Xu, X. He, and X. Yan, Radiation reaction induced spiral attractors in ultra-intense colliding laser beams, Matter Radiat. Extremes 1, 308 (2018).
[78] See Supplemental Material at http://link.aps.org/supplemental/10.1103/PhysRevAccelBeams.22.093401 for three dimensional numerical method and two dimensional PIC simulation results.
[79] T. Kibble, A. Salam, and J. Strathdee, Intensity-dependent mass shift and symmetry breaking, Nucl. Phys. B96, 255 (1975).
[80] C. Harvey, T. Heinzl, A. Ilderton, and M. Marklund, Intensity-Dependent Electron Mass Shift in a Laser Field: Existence, Universality, and Detection, Phys. Rev. Lett. 109, 100402 (2012).
[81] T. G. Blackburn, D. Seipt, S. S. Bulanov, and M. Marklund, Benchmarking semiclassical approaches to strong-field QED: Nonlinear Compton scattering in intense laser pulses, Phys. Plasmas 25, 083108 (2018).
[82] A. Di Piazza, M. Tamburini, S. Meuren, and C. H. Keitel, Implementing nonlinear Compton scattering beyond the local-constant-field approximation, Phys. Rev. A 98, 012134 (2018).
[83] A. Di Piazza, M. Tamburini, S. Meuren, and C. H. Keitel, Improved local-constant-field approximation for strong-field QED codes, Phys. Rev. A 99, 022125 (2019).
[84] V. Dinu, C. Harvey, A. Ilderton, M. Marklund, and G. Torgrimsson, Quantum Radiation Reaction: From Interference to Incoherence, Phys. Rev. Lett. 116, 044801 (2016).
[85] Z. Gong, A. P. L. Robinson, X. Q. Yan, and A. V. Arefiev, Highly collimated electron acceleration by longitudinal laser fields in a hollow-core target, Plasma Phys. Controlled Fusion 61, 035012 (2019).
[86] J. Koga, T. Z. Esirkepov, and S. V. Bulanov, Nonlinear Thomson scattering in the strong radiation damping regime, Phys. Plasmas 12, 093106 (2005).
[87] J. G. Kirk, A. Bell, and I. Arka, Pair production in counter-propagating laser beams, Plasma Phys. Controlled Fusion 51, 085008 (2009).
[88] N. Neitz and A. Di Piazza, Stochasticity Effects in Quantum Radiation Reaction, Phys. Rev. Lett. 111, 054802 (2013).
[89] S. R. Yoffe, Y. Kravets, A. Noble, and D. A. Jaroszynski, Longitudinal and transverse cooling of relativistic electron beams in intense laser pulses, New J. Phys. 17, 053025 (2015).
[90] M. Vranic, T. Grismayer, R. A. Fonseca, and L. O. Silva, Quantum radiation reaction in head-on laser-electron beam interaction, New J. Phys. 18, 073035 (2016).
[91] T. N. Wistisen, A. Piazza, H. V. Knudsen, and U. I. Uggerhøj, Experimental evidence of quantum radiation reaction in aligned crystals, Nat. Commun. 9, 795 (2018).
[92] J. Zuo, M. McCartney, and J. Spence, Performance of imaging plates for electron recording, Ultramicroscopy 66, 35 (1996).
[93] H. Chen, N. L. Back, T. Bartal, F. Beg, D. C. Eder, A. J. Link, A. G. MacPhee, Y. Ping, P. M. Song, A. Throop et al., Absolute calibration of image plates for electrons at energy between 100 KeV and 4 MeV, Rev. Sci. Instrum. 79, 033301 (2008).
[94] B. Pollock, C. Clayton, J. Ralph, F. Albert, A. Davidson, L. Divol, C. Filip, S. Glenzer, K. Herpoldt, W. Lu et al., Demonstration of a Narrow Energy Spread, ∼0.5 Gev Electron Beam from a Two-Stage Laser Wakefield Accelerator, Phys. Rev. Lett. 107, 045001 (2011).